\documentclass[prx,aps,twocolumn,superscriptaddress]{revtex4}

\usepackage{dcolumn}
\usepackage{bm}
\usepackage{amssymb}
\usepackage{color}
\usepackage[pdftex]{graphicx}

\usepackage{amsmath}
\usepackage[linesnumbered,ruled]{algorithm2e}

\usepackage{url}
\usepackage[section]{placeins}
\usepackage[colorlinks=true,linkcolor=blue,citecolor=blue]{hyperref}

\begin{document}


\title{Dynamics of Electronically Phase Separated States in the Double Exchange Model}

\author{Jing Luo}
\affiliation{Department of Physics, University of Virginia, Charlottesville, VA 22904, USA}

\author{Gia-Wei Chern}
\affiliation{Department of Physics, University of Virginia, Charlottesville, VA 22904, USA}

\date{\today}

\begin{abstract}
We present extensive large-scale dynamical simulations of phase-separated states in the double exchange model. These inhomogeneous electronic states that play a crucial role in the colossal magnetoresistance phenomenon are composed of ferromagnetic metallic clusters embedded in an antiferromagnetic insulating matrix. We compute the dynamical structure factor of these nanoscale textures using an efficient real-space formulation of coupled spin and electron dynamics. Dynamical signatures of the various underlying magnetic structures are identified. At small hole doping, the structure factor exhibits a dominating signal of magnons from the background N\'eel order and localized modes from magnetic polarons. A low-energy continuum due to large-size ferromagnetic clusters emerges at higher doping levels. Implications for experiments on magnetoresistive manganites are discussed. 
\end{abstract}

\maketitle

\section{introduction}

Phase separation is ubiquitous in systems dominated by nonlinear and nonequilibrium processes~\cite{seul95,gollub99}. In particular, it has been observed in the intermediate state of numerous first-order phase transitions~\cite{gunton83}. Nanoscale phase separation also underpins many of the intriguing functionalities of strongly correlated electron materials~\cite{dagotto01,moreo99,dagotto05,mathur03,kivelson03,kivelson98}. 
A prominent example is the colossal magnetoresistance (CMR) phenomenon observed in several manganese oxides~\cite{dagotto_book,dagotto01,mathur03,moreo99,salamon01}, in which a small change in magnetic field induces an enormous variation of resistance. Detailed microscopic studies have revealed complex nano-scale textures consisting of metallic ferromagnetic clusters embedded in an insulating matrix~\cite{fath99,renner02,zhang02}. It is believed that CMR arises from a field-induced percolating transition of the metallic nano-clusters in such mixed-phase states~\cite{uehara99,schiffer95,burgy01}. 

Considerable experimental and theoretical effort has been devoted to understanding the origin of these complex mesoscopic textures in manganites and other correlated systems. An emerging picture is that such inhomogeneous states result from the competition between two distinct electronic phases with nearly degenerate energies~\cite{mathur03,dagotto_book}. Microscopically, the double-exchange interaction~\cite{zener51,anderson55,degennes60} is considered a major mechanism for electronic phase separation. The double-exchange model describes itinerant electrons interacting with local magnetic moments through the Hund's rule coupling. Since electrons can gain kinetic energy when propagating in a sea of parallel spins, an instability occurs when ferromagnetic domains favored by doped carriers compete with the background antiferromagnetic order. 
The tendency toward phase separation is further enhanced by factors such as long-range Coulomb interaction, quenched disorder, and coupling to orbital, and lattice degrees of freedom~\cite{dagotto_book}.

The magnetization dynamics of the hole-doped manganites $L_{1-x}A_x$MnO$_3$, where $L$ is a trivalent lanthanide ion and $A$ is a divalent alkaline earth ion, has also been extensively studied experimentally~\cite{zhang07,hwang98,baca98,dai00,ye06,helton17}. The majority of the investigations focused on the ferromagnetic phase with optimal hole doping, which is also the regime exhibiting pronounced CMR effect. While the spin-wave spectrum of some ferromagnetic manganites such as La$_{1-x}$Sr$_x$MnO$_3$ seems well described by the double-exchange model~\cite{furukawa96,motome03}, intriguing unconventional magnetic behaviors have also been reported. For example, the spin wave dispersion of manganites with a lower critical temperature is significantly softened near the Brillouin zone boundary.
Moreover, the magnon excitations close to zone boundary also exhibit an enhanced broadening.
Theoretically, the anomalous spin-wave excitations have been attributed to a host of diverse mechanisms including higher-order effects of spin-charge coupling~\cite{golosov00,shannon02,kapetanakis07}, magnon-phonon interaction~\cite{furukawa99,woods01}, orbital fluctuations~\cite{krivenko04}, and disorder effect~\cite{motome05}.

Despite extensive studies on the spin-wave excitations of the ferromagnetic regime, the spin dynamics of the phase-separated states has received much less attention. It remains poorly understood how the resultant spatial inhomogeneity affects the elementary excitations, which intrinsically involve strongly coupled spin and electron degrees of freedom. Important issues such as the spectrum of coupled electron-magnon dynamics in the phase-separated states have yet to be studied theoretically. The difficulty is partly due to the lack of efficient numerical methods. In particular, the absence of translation invariance in a mixed-phase state renders most momentum-based techniques inapplicable. For example, although the magnon spectrum of the ferromagnetic phase has been computed using generalized spin-wave theories that include the electron-spin interaction~\cite{furukawa96,motome03,golosov00,shannon02,kapetanakis07}, these momentum-based calculations cannot be applied to the phase-separated states of the DE model.

In order to properly account for the spatial inhomogeneity, effective classical or semi-classical spin models, in which the electron degrees of freedom were integrated out beforehand, were employed to compute the magnetic excitation spectrum~\cite{andrade12,drees14}. However, the subtle interplay between the quantum electron degrees of freedom and the spin-dynamics, even treated classical, cannot be captured in such empirical approach. To more properly include the electron effects, an adiabatic dynamical scheme, which integrates out the electrons on the fly of the simulations, have been used to study the dynamical phenomena in inhomogeneous states~\cite{bhattacharyya19,bhattacharyya20}. These approaches, similar to the Born-Oppenheimer approximation in quantum molecular dynamics~\cite{marx09}, assume a fast relaxation of the electron gas. Since the electrons are assumed to remain in the instantaneous {\em equilibrium} states, the adiabatic dynamics not only cannot capture phenomena involving fast electron dynamics, but also fails to describe dynamical phenomena with out-of-equilibrium electrons. A detailed discussion of adiabatic versus non-adiabatic spin-dynamics, in the context of semiclassical spin-density waves, can be found in Ref.~\cite{chern17}. We also discuss the various dynamical modeling for double-exchange-type models in the Appendix.

In this paper, we present the first large-scale {\em non-adiabatic} dynamical simulations of the phase-separated states in the single-band double exchange model based on an efficient real-space method for the entangled dynamics of electrons and spins. In our approach, the time evolution the many-electron wave function, which remains a Slater determinant in double-exchange model, is described by a von~Neumann equation for the reduced density matrix, which is coupled to the Landau-Lifshitz dynamics for spins. By starting from various initial states that represent the thermal ensemble of the system, trajectories of spins are obtained by numerically integrating the coupled spin-electron dynamical equations. 
 The dynamical structure factor of the highly inhomogeneous phase-separated states is then computed via the space-time Fourier transform of the spin trajectories.   Our results reveal intriguing coexistence of ferromagnetic and antiferromagnetic magnons at large hole doping. 
Dynamical signatures of magnetic polarons and ferromagnetic metallic clusters are also identified. In particular, an abundance of low-energy magnons is found to arise from the metallic clusters.

The rest of the paper is organized as follows. In Section~\ref{sec:model} we discuss the single-band double-exchange (DE) model and its various low temperature phases. Numerical methods used to simulate the equilibrium phases of the DE model are also discussed, with an emphasize on the linear-scaling quantum Langevin dynamics method. We next present in Sec.~\ref{sec:dynamics} our {\em non-adiabatic} spin-electron dynamics method. In this framework, the evolution of the DE system is governed by the Landau-Lifshitz equation for classical spins, coupled to the von~Neumann equation for electron correlation function or reduced density matrix. In Sec.~\ref{sec:results}, we apply our non-adiabatic dynamics to compute the dynamical structure factor of electronically phase-separated states in the DE model. Analysis of dynamical signatures associated with magnetic polarons, metallic clusters, and background N\'eel order are also presented. We summarize our work and discuss future applications of our methods in Sec.~\ref{sec:summary}. A detailed account of the various dynamical modeling, including the Langevin thermalization method, adiabatic as well as non-adiabatic Landau-Lifshitz dynamics, is given in the Appendix.


\section{double-exchange model and Langevin dynamics}

\label{sec:model}

We consider the square-lattice double-exchange (DE) model~\cite{zener51,anderson55,degennes60}, which is one of the most representative models that exhibit spontaneous electronic phase separation.  The Hamiltonian of single-band DE model reads
\begin{eqnarray}
	\label{eq:H_DE}
	\hat{\mathcal{H}} = -t \sum_{\langle ij \rangle} \left( \hat{c}^{\dagger}_{i \alpha} \hat{c}^{\;}_{j \alpha} + {\rm h.c.} \right)
	- J \sum_{i} \hat{\mathbf S}_i \cdot \hat{c}^{\dagger}_{i\alpha}  {\bm{\sigma}_{\alpha\beta}} \hat{c}^{\;}_{i\beta},
\end{eqnarray}
where repeated indices $\alpha, \beta$ imply summation. The first term describes the electron hopping: $\hat{c}^\dagger_{i \alpha}$ creates an electron with spin $\alpha = \uparrow, \downarrow$ at site $i$,  $ \langle ij \rangle$ indicates the nearest neighbors, $t$ is the electron hopping constant. The second term represents the Hund's rule coupling between electron spin and the local magnetic moments represented by spin-operator $\hat{\mathbf S}_i$. In order to achieve large-scale simulations, we will use either classical or semiclassical approximation for the local spins. 

The equilibrium phases of this square-lattice DE model have been extensively studied theoretically~\cite{yunoki98,dagotto98,chattopadhyay01}. Exactly at half-filling, the local spins develop a long-range N\'eel order in the $T = 0$ insulating ground state. At small electron densities, on the other hand, a metallic state with predominantly ferromagnetic (FM) spin correlation emerges as the ground state. Near half-filling with a small hole doping, the FM metal becomes unstable against either a noncollinear magnetic spiral or phase separation~\cite{yunoki98,dagotto98,chattopadhyay01,azhar17} depending on the strength of the Hund's coupling $J$. 
In the large-$J$ regime, the instability of the FM phase leads to phase separation with coexisting FM and N\'eel domains, as directly verified in Monte Carlo simulations~\cite{yunoki98}.

To obtain the equilibrium phases of the DE system, including the phase-separated states, conventional Monte Carlo simulations are applied to sample the classical spin configurations of the DE Hamiltonian. This is because for a given spin configuration, the DE Hamiltonian, which is quadratic in fermion operators, is essentially a tight-binding model, hence can be exactly solved.  For each spin update $\{\mathbf S_i \} \,\to \, \{\mathbf S_i'\}$, one first computes the free-energy difference $\Delta \mathcal{F} = \mathcal{F}' - \mathcal{F}$, where the free-energy for a given spin configuration is obtained by integrating out the electrons
\begin{eqnarray}
	\mathcal{F}(\{\mathbf S_i\}) &=& -k_B T \,\ln \mathcal{Z}_c \nonumber \\
	& = & -k_B T\,\ln {\rm Tr}_c\, \exp\left[{-\beta \hat{ \mathcal{H}}(\{\mathbf S_i\} )}\right].
\end{eqnarray}
Then the Metropolis formula: $P = {\rm min}\left(1, e^{-\Delta\mathcal{F}/k_B T} \right)$ is used to determine whether the proposed spin-update is accepted. However, Repeated calculation of $\Delta\mathcal{F}$ based on, e.g. exact diagonalization, can be overwhelmingly time consuming. Efficient linear-scaling techniques, such as the kernel-polynomial method (KPM), often have to be used in order to achieve large-scale equilibrium simulations of the DE model~\cite{furukawa04,alvarez05,weisse06}. However, even with KPM, most local spin-update Monte Carlo methods are not very effective. Here we adopt an efficient adiabatic Langevin dynamics method combined with a gradient extension of the KPM~\cite{barros13,barros14,wang18} to obtain equilibrium phase-separated states. In this approach, the time evolution of the classical spins is governed by the stochastic model-A dynamics~\cite{hohenberg77}
\begin{eqnarray}
	\label{eq:langevin}
	\frac{d\mathbf S_i}{dt} = -\alpha \frac{\partial \mathcal{ F}}{\partial \mathbf S_i} + \bm\xi_i(t) = \alpha \mathbf T_i + \bm\xi_i(t),
\end{eqnarray}
where $\alpha$ is a damping coefficient, and $\bm\xi_i(t)$ is a $\delta$-correlated fluctuating force satisfying $\langle \xi_i(t) \rangle = 0$, and $ \langle \xi^\mu_i(t) \xi^\nu_j(t') \rangle = 2 \alpha k_B T \delta_{ij} \delta_{\mu\nu} \delta(t - t')$.
An implicit Lagrangian multiplier is used to enforce the constant-length constraint $|\mathbf S_i(t)| = 1$. Following standard procedure by deriving the associated Fokker-Planck equation, it can be shown that the stead-state configurations are described by the desired Boltzmann distribution.

It should be noted that the Langevin equation Eq.~(\ref{eq:langevin}) describes pure relaxational dynamics of the DE system, which completely ignores the precessional dynamics of spins. In this sense, one cannot apply it to model the general dynamical phenomena.  However, for the purpose of sampling spin-configurations in equilibrium, the Langevin method is sufficient and very effective, especially compared with Monte Carlo method. An alternative approach is the quantum version of the stochastic Landau-Lifshitz-Gilbert dynamics~\cite{chern17}, which incorporates both precession and relaxation spin-dynamics; see Appendix for more details. However, as already noted in the Introduction, such approach is based on the adiabatic approximation. As a result, it cannot be used to describe dynamical process with a time-scale shorter than the electron relaxation times.

\section{Landau-Lifshitz von~Neumann dynamics}

\label{sec:dynamics}

The adiabatic Landau-Lifshitz dynamics, which is similar to the Born-Oppenheimer approximation in quantum molecular dynamics~\cite{marx09}, is based on the assumption that the time-scale of electron motion and that of spin dynamics are well separated. Assuming much faster electronic relaxation, the electrons are assumed to remain in equilibrium for any given spin configuration. Consequently, such dynamical model does not account for the feedback of spin dynamics to the electronic degrees of freedom. More importantly, even without Gilbert damping and stochastic noise, the total energy of the DE system is {\em not} conserved under the adiabatic Landau-Lifshitz dynamics. The adiabatic approximation is thus inadequate for describing the fundamental entangled dynamics of coupled electron-spin system. We note that adiabatic dynamics for spin-density waves has been formulated in Ref.~\cite{chern17}. Moreover,  adiabatic {\em ab initio} spin-dynamics has also been developed based on the adiabatic approximation of the time-dependent density functional theory~\cite{niu99,capelle01,qian02}.

To go beyond the adiabatic approximation, one needs to treat the electron dynamics on an equal footing with the spin dynamics. Here we present a non-adiabatic dynamics method for the DE system, which is similar in spirit to the Ehrenfest method in quantum molecular dynamics simulations~\cite{tully90,li05,marx09}. We assume that the quantum state of the DE system is a direct product state: $|\Omega(t) \rangle = |\Phi(t) \rangle \otimes |\Psi(t)\rangle$, where $|\Phi\rangle$ and $|\Psi\rangle$ denote the spin and electron wave functions, respectively. Consistent with the classical approximation for spins, we assume $|\Phi\rangle$ is a direct product of single-site spin-state: $|\Phi(t)\rangle = \prod_i |\phi_i(t)\rangle$. This is equivalent to a mean-field approximation for the quantum spins. We then define time-dependent local moment $\mathbf S_i(t) = \langle \phi_i(t) | \hat{\mathbf S}_i | \phi(t) \rangle$ as the expectation value of the spin operator.

By tracing out the spin degrees of freedom, the evolution of the electron wavefunction $|\Psi(t)\rangle$ is described by the Schr\"odinger equation with a time-varying DE Hamiltonian
\begin{eqnarray}
	i\hbar \frac{d|\Psi(t) \rangle}{dt} = \hat{\mathcal{H}}_{\rm e}(\{\mathbf S_i(t) \}) |\Psi(t) \rangle.
\end{eqnarray} 
where the effective electron Hamiltonian is defined as
\begin{eqnarray}
	\hat{\mathcal{H}}_e(\{\mathbf S_i \}) = \langle \Phi(t) | \hat{\mathcal{H}} | \Phi(t) \rangle,
\end{eqnarray}
and $\hat{\mathcal{H}}$ is the DE Hamiltonian in Eq.~(\ref{eq:H_DE}). Next, to derive the spin dynamics, we switch to the Heisenberg picture. The equation of motion for vector $\mathbf S_i(t)$ is then
\begin{eqnarray}
	\frac{d \mathbf S_i}{dt} = \frac{i}{\hbar} \left\langle [\hat{\mathcal{H}}, \hat{\mathbf S}_i ] \right\rangle,
\end{eqnarray}
where $\langle \cdots \rangle$ means expectation value computed using the full quantum state $|\Omega\rangle$. Using the product nature of the spin-state $|\Phi\rangle$, the above equation can be expressed as the standard Landau-Lifshitz (LL) equation
\begin{eqnarray}
	\label{eq:LL0}
	\frac{d \mathbf S_i}{dt} =   \mathbf S_i \times \frac{\partial \mathcal{E}}{\partial \mathbf S_i},
\end{eqnarray}
where the effective energy is
\begin{eqnarray} 
	\mathcal{E}(\{\mathbf S_i\}) = \langle \Psi(t) | \hat{\mathcal{H}}_e | \Psi(t) \rangle,
\end{eqnarray} 
It is worth noting that the equivalence between Landau-Lifshitz dynamics and the mean-field type approximation for quantum spin dynamics has been pointed out in earlier works~\cite{fazekas99,ma12}. Since the linearized Landau-Lifshitz equation is equivalent to the leading-order Holstein-Primarkoff expansion, the above treatment can also be viewed as a semiclassical dynamics for  quantum spins in the DE model.  We also note that, classical spin approximation is widely used in the study of CMR phenomena. Our approach here thus corresponds to the dynamical generalization of the classical picture used for DE-type systems.

Substituting the explicit form of the DE Hamiltonian~Eq.~(\ref{eq:H_DE}) into the above Landau-Lifshitz equation, we obtain
\begin{eqnarray}
	\label{eq:LL}
	\frac{d \mathbf S_i}{dt}  = - J \mathbf S_i \times \bm{\sigma}^{\;}_{\alpha\beta}\,\, \rho^{\;}_{i\beta,i\alpha}(t).
\end{eqnarray}
Here we have have defined the time-dependent single-electron density matrix or correlation function 
\begin{eqnarray}
	\rho^{\;}_{i\alpha, j\beta}(t) \equiv \langle \Psi(t) | \hat{c}^\dagger_{j\beta}\,\hat{c}^{\;}_{i\alpha} |\Psi(t) \rangle.
\end{eqnarray} 
Since the DE Hamiltonian is bilinear in fermion operators, the many-electron wave function $|\Psi(t) \rangle$ remains a single Slater-determinant state if it is a pure state initially. However, the integration of the Slater determinant is rather cumbersome numerically and has only been used for rather small size simulations~\cite{koshibae11,ono17}.

Instead of evolving the Slater determinant wavefunction, one could equally describe the electron dynamics in terms of the reduced density matrix based on the Heisenberg picture. An additional advantage of this formulation is that it can account for the initially mixed quantum states. Detailed of the generalization to include the ensemble average of the initial states can be found in Appendix. 
To describe the electron dynamics in our formalism, we define a time-dependent ``first-quantization" Hamiltonian matrix 
\begin{eqnarray}
	\label{eq:H_matrix}
	H_{i\alpha, j\beta} = -t_{ij} \delta_{\alpha\beta} - J \mathbf S_i \cdot \bm \sigma_{\alpha\beta} \,\delta_{ij},
\end{eqnarray}
The DE model in Eq.~(\ref{eq:H_DE}) can then be expressed as 
\begin{eqnarray}
 	\hat{\mathcal{H}} =\sum_{i,j} \sum_{\alpha, \beta} H_{i\alpha, j\beta}  \hat{c}^{\dagger}_{i \alpha} \hat{c}^{\;}_{j \beta}.
\end{eqnarray} 
In terms of matrix $H$, the reduced density matrix satisfies the von~Neumann equation 
\begin{eqnarray}
	\frac{d\rho}{dt} = i [\rho, H]. 
\end{eqnarray}	
Substituting Eq.~(\ref{eq:H_matrix}) for $H$ to the above equation yields
\begin{eqnarray}
	\label{eq:vN}
	\frac{d\rho_{i\alpha, j\beta}}{dt} &=& i (t_{ik} \, \rho_{k\alpha, j\beta} - \rho_{i\alpha, k\beta} \, t_{kj})  \\
	&+& i J (\mathbf S_i \cdot \bm \sigma_{\alpha\gamma} \, \rho_{i\gamma, j\beta} - \rho_{i\alpha, j\gamma} \, \bm\sigma_{\gamma\beta} \cdot \mathbf S_j). \nonumber
\end{eqnarray}
It can be readily verified that the total energy of the system $E = \langle \mathcal{H} \rangle = {\rm Tr}(\rho H)$ is a constant of motion.
The numerical efficiency of integrating the von~Neumann equation can be improved with optimized sparse-matrix multiplication algorithms. A similar formulation has been developed for the semiclassical dynamics of spin density waves in the Hubbard model~\cite{chern17}.

\section{dynamical structure factor of phase-separated states}

\label{sec:results}

\begin{figure*}
\includegraphics[width = 2\columnwidth]{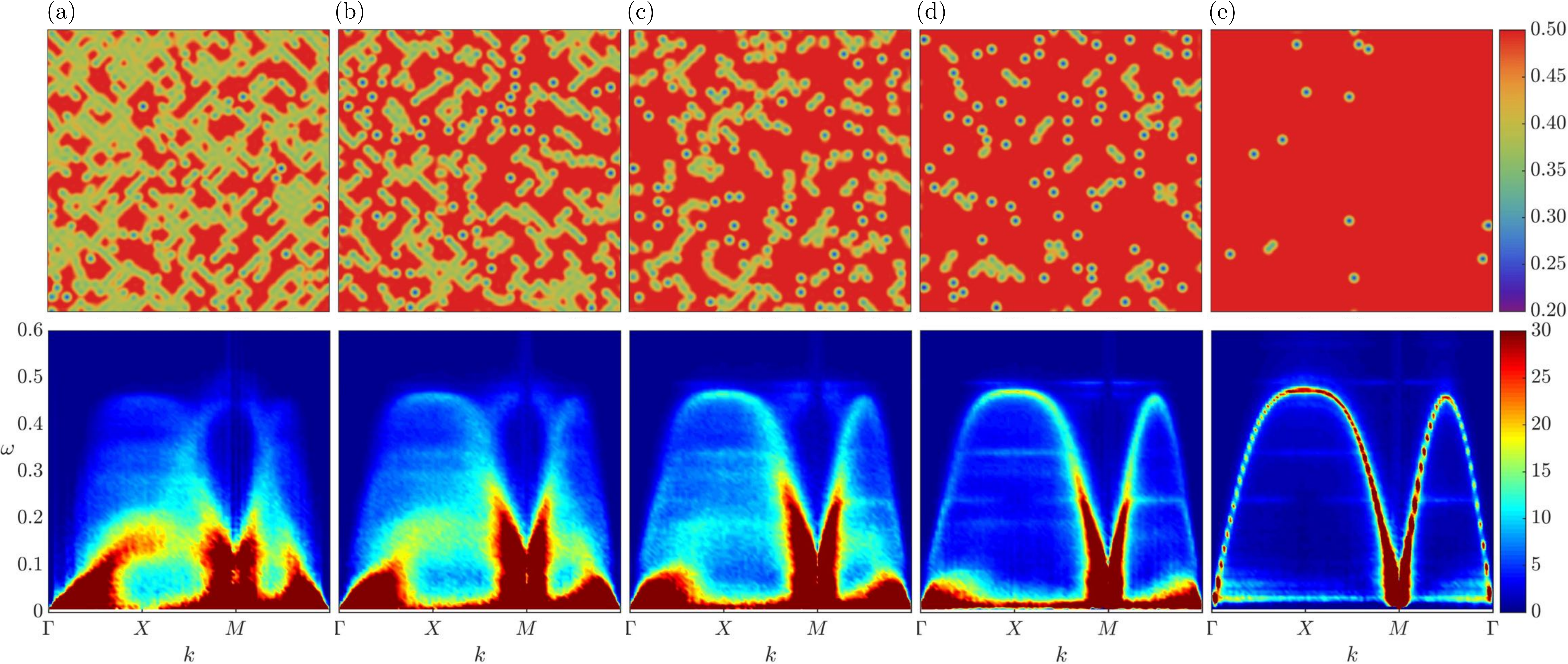}
\caption{
\label{fig:phase-DSF}
Upper panels: density plots of the on-site electron number $n(\mathbf r_i) = \langle \hat{c}^\dagger_{i,\alpha} \hat{c}^{\;}_{i, \alpha} \rangle$ in sample phase-separated states for varying electron filling fraction $f = \sum_i n(\mathbf r_i)/2N$, where $N$ is the number of lattice sites, obtained from Langevin dynamics simulations on a $60\times 60$ lattice with Hund's coupling $J = 6t$ and temperature $T = 5 \times 10^{-4} t$. The corresponding dynamical structure factors $\mathcal{S}(\mathbf q, \omega)$ averaged over tens of independent initial states are shown in the lower panels. The high-symmetry points of the Brillouin zone are $\Gamma=(0,0), X=(\pi,0), M=(\pi,\pi)$.}
\end{figure*}

To compute the dynamical structure factor of the phase-separated states, the initial state is prepared using the KPM Langevin simulations at a temperature of $T = 5\times 10^{-4} \,t$. A few examples of such mixed-phase states on a $60\times 60$ square lattice are shown in the upper panels of Fig.~\ref{fig:phase-DSF}. The red region corresponds to the half-filled insulating background with the antiferromagnetic order, while the green and blue regions indicate metallic FM domains with low electron density. Interestingly, in addition to forming the FM puddles, a fraction of the doped holes are self-trapped in a composite object which can be viewed as the magnetic analog of  polaron~\cite{furuwaka94,varma96,coey95,yi00}. 

For each of the initial configuration prepared by the Langevin simulations, a fourth-order Runge-Kutta method is used to integrate the non-adiabatic Landau-Lifshitz-von~Neumann equations, i.e. Eqs~(\ref{eq:LL}) and (\ref{eq:vN}). From the numerically obtained spin trajectories $\mathbf S_i(t)$, we compute the dynamical correlation function  
\begin{eqnarray}
	\mathcal{S}(\mathbf q, t) = \langle \mathbf S_{\mathbf q}(t) \cdot \mathbf S_{\mathbf q}^*(0) \rangle,
\end{eqnarray}
where $\mathbf S_{\mathbf q}(t) \equiv \sum_i \mathbf S_i(t) \exp(i \mathbf q \cdot \mathbf r_i) / \sqrt{N}$ is the spatial Fourier transform of the instantaneous spin configuration, and $\langle \cdots \rangle$ denotes the ensemble average over independent initial states of a given temperature. The dynamical structure factor is then given by
\begin{eqnarray}
	\mathcal{S}(\mathbf q, \omega) &=& \int \mathcal{S}(\mathbf q, t) e^{-i \omega t} dt \nonumber \\
	& = & \frac{1}{N}\sum_{ij} \int dt \langle \mathbf S_i(t) \cdot \mathbf S_j(0) \rangle e^{-i \omega t} dt,
\end{eqnarray}
which is essentially the space-time Fourier transform of the spin-spin correlator   $C(\mathbf r_{ij}, t) = \langle \mathbf S_i(t) \cdot \mathbf S_j(0) \rangle$. Importantly, the dynamical simulation here is completely deterministic and energy-conserving.  The lower panels of Fig.~\ref{fig:phase-DSF} show the $\mathcal{S}(\mathbf q, \omega)$ of phase-separated states with five different electron filling fractions; each is averaged over 50 distinct initial states. 

Since the N\'eel order parameter, characterized by the wavevector $\mathbf Q = (\pi, \pi)$ at the $M$-point, is not a conserved quantity, the fluctuations of the associated Fourier component $\tilde S(\mathbf Q, t) \equiv \sum_i \mathbf S_i(t) \, e^{i \mathbf Q \cdot \mathbf r_i}$ produce a huge artifact in the raw data of the dynamical structure factor. Interestingly, we found that the drifting of this Goldstone mode of finite lattices exhibits a $1/\omega$ power-law behavior, extending to very high energies. This observation thus allows us to systematically remove the large artificial signal in the vicinity of the $M$-point. The $\mathcal{S}(\mathbf q, \omega)$ shown in Fig.~\ref{fig:phase-DSF} were obtained after this subtraction.

\begin{figure}[b]
\includegraphics[width = 0.99\columnwidth]{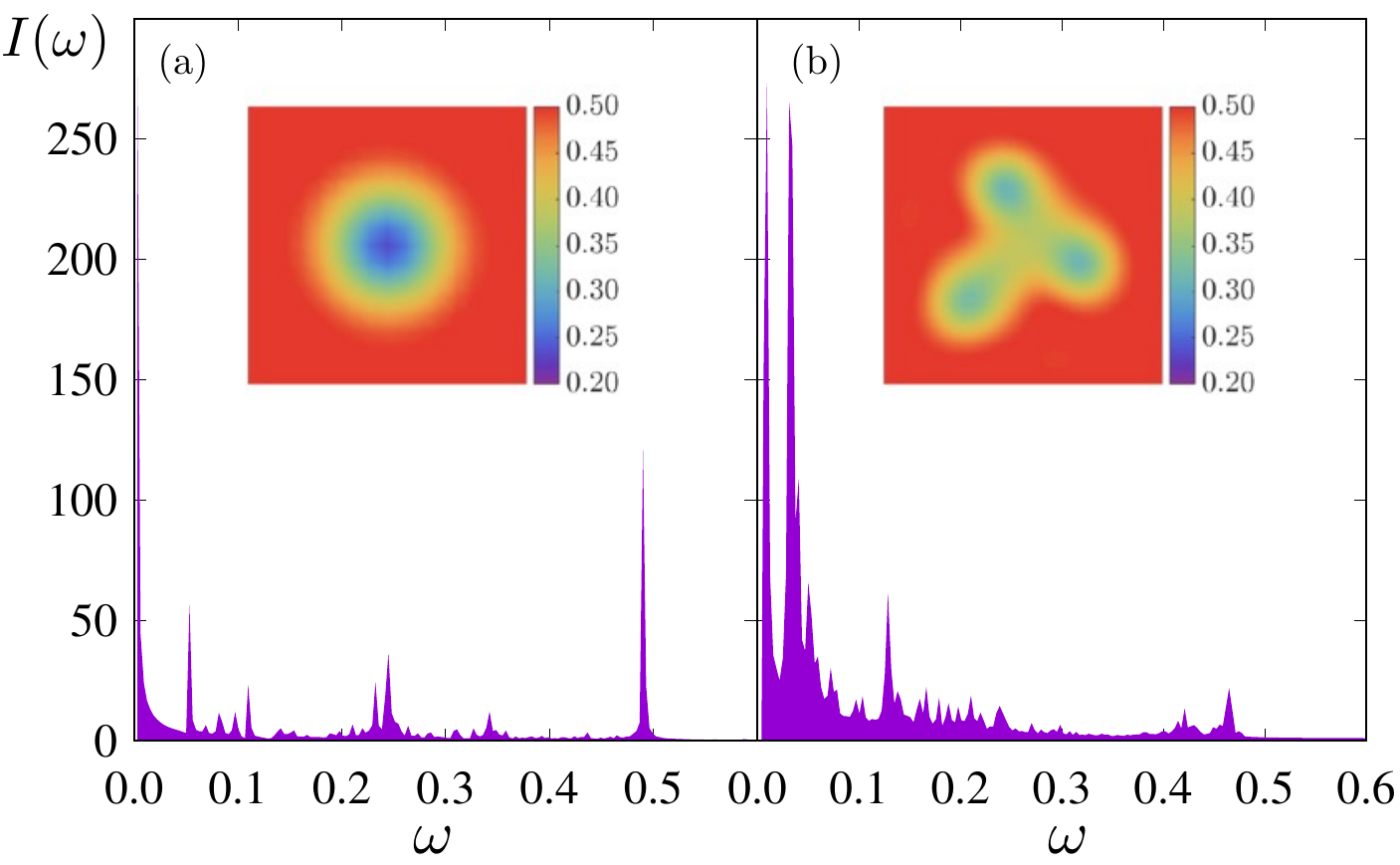}
\caption{
\label{fig:spectrum}
The power spectrum $I(\omega) = \sum_{i \in \mathcal{C}} |\tilde{\mathbf S}_i(\omega)|^2$ of (a)~a magnetic polaron and (b) a FM metallic cluster consisting of roughly 20 spins. Here the sum runs over spins in the FM domain of these object. The inset shows the electron density plot $n(\mathbf r_i) = \langle c^\dagger_{i \alpha} c^{\;}_{i \alpha} \rangle$. With a large Hund's coupling $J = 6t$, the size of the magnetic polaron is rather small, with a radius of roughly three lattice constants. 
}
\end{figure}

The dynamical structure factor in the vicinity of half-filling is dominated by the background antiferromagentic spin-wave excitations, as shown in Fig.~\ref{fig:phase-DSF}(e). The pronounced signals around the $M$-point correspond to the Goldstone modes of the underlying N\'eel order. As mentioned above, the doped holes in this regime are localized by the self-induced potential in a magnetic polaron. Numerically, each polaron is found to accommodate nearly exactly one hole. To understand the nature of the associated spin excitations, we focus on the dynamics of a single magnetic polaron. We first perform relaxational dynamics on a perturbed half-filled N\'eel state (by flipping a center spin) with exactly one electron removed to obtain the initial states. From the spin dynamics, we compute the power spectrum $I(\omega) \equiv \sum_{i \in \mathcal{C}} | \tilde{\mathbf S}_i(\omega) |^2$, where $\tilde{\mathbf S}_i(\omega)=\int \mathbf S_i(t) e^{-i\omega t} dt$ and the summation is over five spins at the center of the polaron. The computed spectrum, shown in Fig.~\ref{fig:spectrum}(a), is characterized by prominent peaks at, e.g. $\omega/t = 0.05, 0.25, 0.49$, corresponding to eigen-energies of the spin-wave excitations localized at the magnetic polaron. Importantly, these localized magnons contribute to the flat bands seen in the $\mathcal{S}(\mathbf q, \omega)$.

With increasing hole doping, the antiferromagnetic spin-wave dispersion is still visible, yet with gradually reduced strength. Some of the flat-bands due to magnetic polarons also persist. An intriguing new feature is the emergence of a continuum of low-energy magnons throughout the whole Brillouin zone. It is tempting to associate this continuum with the metallic FM clusters whose size also grows with increasing hole doping; see Fig.~\ref{fig:phase-DSF}. To this end, we examine the spectrum of metallic clusters of varying shapes and sizes. Similar to the preparation of the magnetic polaron, we manually create such structures by carefully tuning the hole doping with the cluster size. Fig.~\ref{fig:spectrum}(b) shows the $I(\omega)$ of a sample cluster consisting of roughly 20 spins. A few pronounced peaks, corresponding to the dominant quantized magnons, can be seen in the spectrum. While the intensity and position of these peaks depend on the geometric details of the FM clusters, a common feature of the cluster spectrum is the appearance of numerous low energy modes.

\begin{figure}[t]
\includegraphics[width = 1\columnwidth]{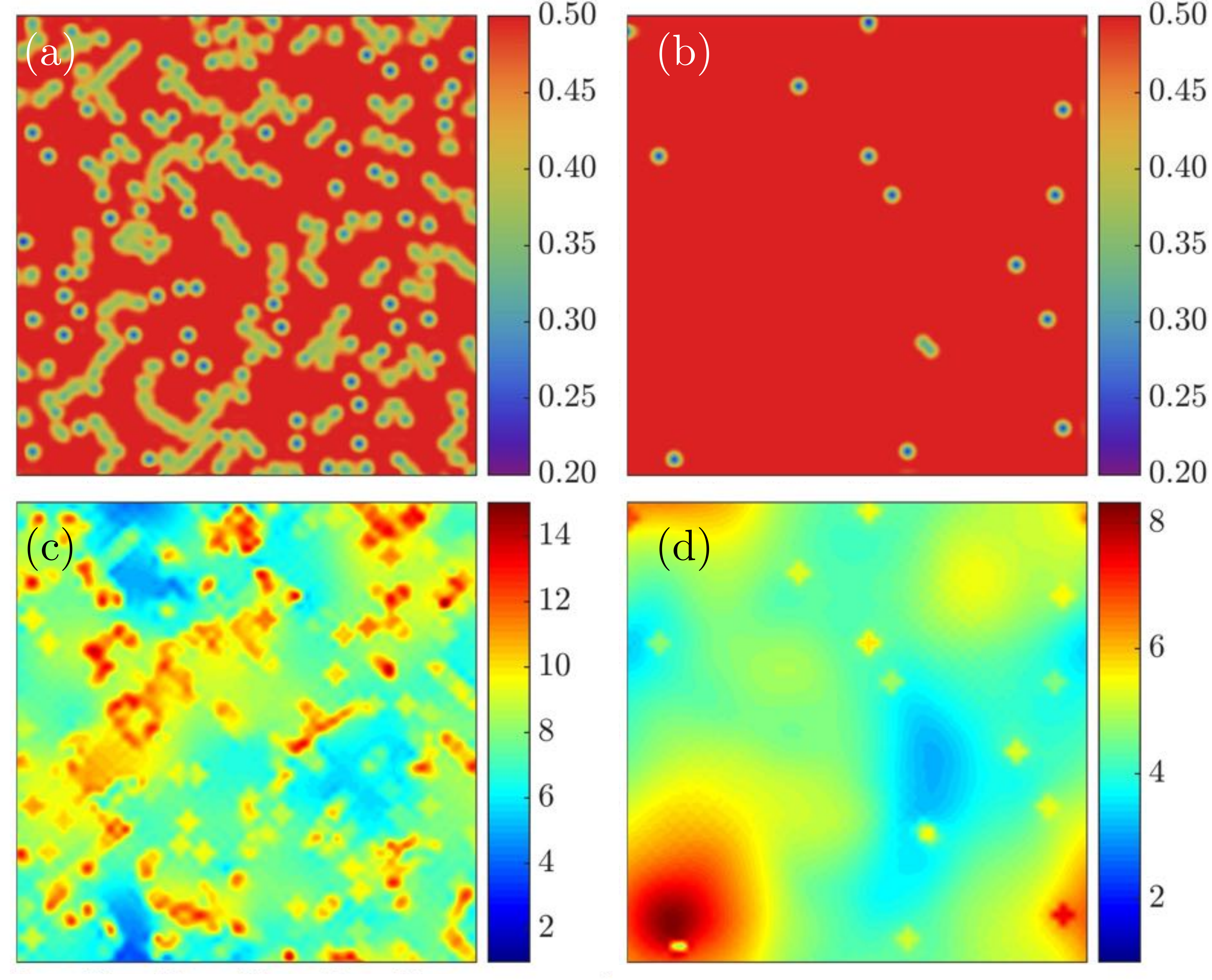}
\caption{
\label{fig:profile}
Top panels show the density plot $n(\mathbf r_i) = \langle c^\dagger_{i\alpha} c^{\;}_{i\alpha} \rangle$ of one particular phase-separated state for filling fractions (a)~$f = 0.465$ and (b) $f = 0.498$. The simulated system size is $60\times 60$. The corresponding spatial profile of spin excitations  $\mathcal{F}(\mathbf r_i) = \int_{\omega_1}^{\omega_2} |\tilde{\mathbf S}_i(\omega)|^2\, d\omega$ is shown in panels (c) and (d), respectively, where $\omega_1 = 0.006283$ and $\omega_2 =  0.09425$. 
}
\end{figure}

To further investigate the nature of these low-energy magnons, we compare their spatial profile $\mathcal{F}(\mathbf r)$ with the corresponding electron density plot $n(\mathbf r)$ for a particular initial state, as demonstrated in Fig.~\ref{fig:profile} for two electron filling fractions. Here the magnon profile function is defined as the integral of the spin Fourier components $\mathcal{F}(\mathbf r_i) = \int_{\omega_1}^{\omega_2} |\tilde{\mathbf S}_i(\omega)|^2\, d\omega$, over a finite band $[\omega_1, \omega_2]$ of small energies. In the case of electron filling $n = 0.465$, where the system is spontaneously segregated into FM domains of various sizes in an AFM background, the dominant spin excitations in this energy range are from the FM clusters of the doped holes; see Fig.~\ref{fig:profile}(a) and~(c). For even smaller hole doping with $f = 0.498$, the intensity plot of $\mathcal{F}(\mathbf r)$ exhibits a complex long-wavelength pattern of the N\'eel background as shown in Fig.~\ref{fig:profile}(d). Distinctive signals can be seen that are contributed from the small-size magnetic polarons. 

\begin{figure}[t]
\includegraphics[width = 0.99\columnwidth]{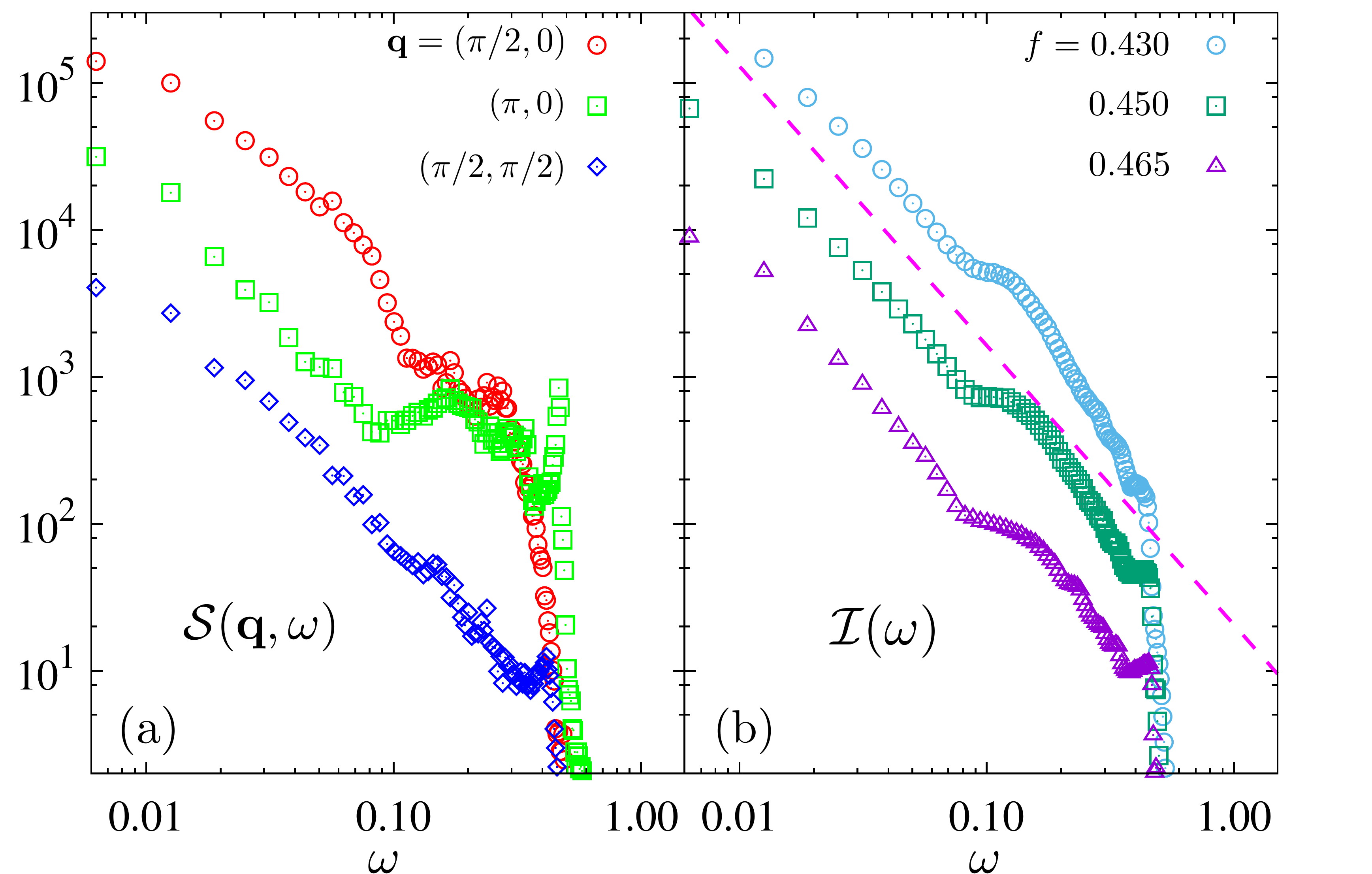}
\caption{
\label{fig:omegaplot}
(a) The dynamical structure factor $\mathcal{S}(\mathbf q, \omega)$ versus $\omega$ at a few selected wavevectors $\mathbf q$ for filling fraction $f = 0.465$. (b) The frequency dependence of the spin excitation spectrum $\mathcal{I}(\omega) \equiv \sum_{\mathbf q} \mathcal{S}(\mathbf q, \omega) / N$ integrated over the whole Brillouin zone, at varying electron filling fractions. The curves are shifted vertically for clarity. The dashed line shows the $\omega^{-2.1}$ power-law dependence.
}
\end{figure}

We next examine the spectral distribution of low-energy spin excitations of the phase-separated states. Fig.~\ref{fig:omegaplot}(a) shows the log-log plot of the dynamical structure factor $\mathcal{S}(\mathbf q, \omega)$ versus $\omega$ at a few selected wavevectors. Each curve is again obtained after averaging over tens of different mixed-phase configurations. These distributions exhibit an abrupt drop above a band-edge $E_b \sim 0.5t$, indicating the absence of magnon density of states at high energies; see also the density plots in Fig.~\ref{fig:phase-DSF}. While the dynamical structure factor at the three different $\mathbf q$'s shows rather distinct $\omega$ dependences at high energies, a pronounced increase of $\mathcal{S}(\mathbf q, \omega)$ at small $\omega$ can be seen for all three wavevectors shown in Fig.~\ref{fig:omegaplot}(a).

The overall density of states (DOS) of the low-energy magnons can be inferred from the spectral function $\mathcal{I}(\omega) = \sum_{\mathbf q} \mathcal{S}(\mathbf q, \omega) / N$, which is the dynamical structure factor averaged over the whole Brillouin zone. Fig.~\ref{fig:omegaplot}(b) shows the log-log plots of the numerical spectral function $\mathcal{I}(\omega)$ for three different filling fractions. Interestingly, they show strong similarities with each other, especially with increasing hole doping. At high energies, one can see a clear band-edge and a shoulder-like feature. The distribution functions $\mathcal{I}(\omega)$ develop a sharp peak at $\omega \to 0$, which indicates a significant increase in the magnon DOS at small $\omega$. The nearly linear segments in the log-log plot of Fig.~\ref{fig:omegaplot}(b) suggest a power-law behavior.  It is worth noting that a similar disorder-induced peak at  $\omega \to 0$ also appears in the magnon DOS in different localized spin models~\cite{chakraborty10,vojta13}. 

In our case, these low-energy magnons can be viewed as descending from the zero-energy modes of individual isolated FM clusters. These zero modes acquire a finite energy through coupling to the N\'eel background. For short-range spin-spin interactions, this energy shift of the zero modes is expected to scale as the circumference of the FM cluster. Since the electron-mediated spin interactions are long-ranged, the energy shift might scale differently. For simplicity, we assume a power-law relation $\omega \sim s^{1/\Delta}$ between the acquired energy $\omega$ of the zero mode and the size $s$ of the FM puddle. For example, $\Delta = 2$ corresponds to the case $\omega \sim \ell$, where $\ell$ is the linear cluster size. The density of states $\rho(\omega)$ is then related to the distribution of $s$ through $\rho(\omega) \sim n(s)\, \omega^{\Delta-1}$. In the vicinity of the cluster percolation transition, one expects a power-law cluster-size distribution $n(s) \sim 1/s^\tau$ for large clusters; here $\tau$ is the Fisher's exponent~\cite{stauffer94}. This in turns indicates a power-law DOS $\rho(\omega) \sim 1/\omega^{\Delta \, (\tau -1) + 1}$. Numerically, we find that $\Delta$ is very close to 1~\cite{note_tau}.


\section{summary and outlook}

\label{sec:summary}

To summarize, we have presented an efficient numerical framework for the real-space dynamical simulation of the double-exchange model. Focusing on the regime with small hole doping, we compute, for the first time, the dynamical structure factor of the electronically phase-separated states. In particular, we found an emerging low-energy magnon continuum that is attributed to the quasi-zero-modes of large-size metallic FM clusters. Such abundance of low-energy magnons has been observed in recent neutron-scattering measurement of ferromagnetic manganites La$_{0.7}$Ca$_{0.3}$MnO$_3$ close to optimal doping~\cite{helton17}. We have also observed the coexistence of FM and AFM magnons in the dynamical structure factor with larger hole doping, a result consistent with recent experiment~\cite{chatterji06}.  Our work thus opens a new avenue for dynamical simulations of electronic inhomogeneous states in other systems.



It is worth pointing out that magnon spectrum in real materials also depends on factors such as long-range dipole-dipole or Coulomb interactions, anisotropy, charge-phonon, and charge-orbital coupling. Our goal here is to examine the intrinsic effects of phase separation on the coupled electron-spin dynamics.  While several mechanisms, such as magnon-phonon interaction~\cite{furukawa99,woods01}, orbital fluctuations~\cite{krivenko04}, and disorder effect~\cite{motome05}, have been proposed to explain the unusual softening and broadening of spin-waves in some manganites, recent experiment~\cite{helton17} highlighted the possibility that this anomalous behavior could be simply caused by the electronic phase separation. With our efficient formulation, it is now possible to quantitatively study the FM magnons in the mixed-phase state, which will be left for future study.

\bigskip

\begin{acknowledgments}
We thank Kipton Barros for the help with the kernel polynomial method. This work is partially supported by the Center for Materials Theory as a part of the Computational Materials Science (CMS) program, funded by the US Department of Energy, Office of Science, Basic Energy Sciences, Materials Sciences and Engineering Division. The authors also acknowledge the support of Advanced Research Computing Services at the University of Virginia.
\end{acknowledgments}

\appendix*

\section{Dynamical modeling for double-exchange systems}

\label{app:dyn}

Most numerical investigations of the double-exchange (DE) systems~\cite{zener51,anderson55,degennes60} have so far focused on their equilibrium properties. In particular, extensive Monte Carlo simulations and dynamical mean-field theory calculations have been used to obtain the phase diagrams. However, dynamical simulations of inhomogeneous phases in DE models have yet to be explored.  In this supplemental material, we discuss the various dynamics of the double-exchange (DE) system and their numerical formulations. We first review the adiabatic spin dynamics, which is analogous to the Born-Oppenheimer approximation in quantum molecular dynamics simulations~\cite{marx09}. Next we discuss the non-adiabatic Ehrenfest dynamics for coupled electron-spin systems. Finally we present details of the Landau-Lifshitz-von~Neumann method for the non-adiabatic dynamics of the DE system.

\subsection{Adiabatic spin dynamics}

In the adiabatic approximation, which is similar to Born-Openheimer approximation for quantum molecular dynamics~\cite{marx09}, electrons are assumed to quickly relax to the equilibrium state corresponding to the instantaneous spin configuration $\{\mathbf S_i \}$. One can thus define an effective energy functional for the spins by integrating out electrons
\begin{eqnarray}
	\mathcal{F}(\{\mathbf S_i\}) = -k_B T \,\ln \mathcal{Z}_c \equiv -k_B T\,\ln {\rm Tr}_c\left(e^{-\beta \mathcal{H}}\right),  
\end{eqnarray}
which is essentially the canonical free energy of the electrons. Here $\beta = 1/k_B T$ is the inverse temperature, and we have also defined the electron partition function $\mathcal{Z}_c$. This free energy can also be written as
\begin{eqnarray}
	\label{eq:free-energy}
	\mathcal{F} = \mathcal{E} - T \mathcal{S}
\end{eqnarray}
where $\mathcal{E}$ is the internal energy
\begin{eqnarray}
	\mathcal{E} = \langle \mathcal{H} \rangle =\frac{1}{\mathcal{Z}_c} {\rm Tr}_{c}\left(e^{-\beta \mathcal{H}} \, \mathcal{H} \right),
\end{eqnarray} 
and $\mathcal{S}$ is the entropy of the electron system. In terms of the eigen-energies $\varepsilon_m = \varepsilon_m(\{\mathbf S_i \})$, which are functions of the instantaneous spin configuration, the electron energy and entropy can be expressed as
\begin{eqnarray}
	\label{eq:E_elec1}
	\mathcal{E}  = \sum_m \varepsilon_m \, f_m , 
\end{eqnarray}
\begin{eqnarray}
	\label{eq:S_elec}
	\mathcal{S} = -k_B \sum_m \left[ f_m \ln f_m + (1 - f_m) \ln(1- f_m) \right],
\end{eqnarray}
where $f_m = 1/\left( \exp(\beta \varepsilon_m) + 1\right)$ is the Fermi-Dirac function for the electron occupation. For simplicity, here we have set the zero of the electron energy to be the chemical potential. 

Given this effective spin energy $\mathcal{F}$, the torques acting on spins can be defined as
\begin{eqnarray}
	\label{eq:torque}
	\mathbf T_i = -\frac{\partial \mathcal{F}}{\partial \mathbf S_i}.
\end{eqnarray}
This torque can also be thought of as an effective local exchange field, which plays a central role in spin dynamics. For example, the pure relaxational dynamics, or model~A~\cite{hohenberg77}, is described by the following stochastic equation
\begin{eqnarray}
	\frac{d\mathbf S_i}{dt} = -\alpha \frac{\partial \mathcal{ F}}{\partial \mathbf S_i} + \bm\xi_i(t) = \alpha \mathbf T_i + \bm\xi_i(t),
\end{eqnarray}
where $\alpha$ is a damping coefficient, and $\bm\xi_i(t)$ is a $\delta$-correlated fluctuating force satisfying
\begin{eqnarray}
	\langle \xi_i(t) \rangle = 0, \,\, \langle \xi^\mu_i(t) \xi^\nu_j(t') \rangle = 2 \alpha k_B T \delta_{ij} \delta_{\mu\nu} \delta(t - t').
\end{eqnarray}
An implicit Lagrangian multiplier is used to enforce the constraint $|\mathbf S_i(t)| = 1$.
This equation can also be viewed as the Langevin dynamics of the spins~\cite{barros13,barros14}, which has proven a powerful method for simulating equilibrium phases of the DE systems.  More generally, the spin dynamics is described by the stochastic Landau-Lifshitz or Landau-Lifshitz-Gilbert (LLG) equation~\cite{brown63,antropov97,ma11}
\begin{eqnarray}
	\label{eq:LLG}
	\frac{d \mathbf S_i}{dt}\ = - \mathbf S_i \times \left( \mathbf T_i + \bm\xi_i \right) - \alpha \mathbf S_i \times \left( \mathbf S_i \times  \mathbf T_i \right).
\end{eqnarray}
Here the first term on the right hand side describes the energy-conserving precessional dynamics of spins, while the second term represents the dissipation effect. This stochastic LLG equation can be used to simulate both equilibrium or non-equilibrium magnetic dynamical phenomena. However, it should be noted that electrons are assumed to be in quasi-equilibrium state within the adiabatic approximation.

The most time-consuming part of such adiabatic dynamical simulations, either the relaxational or the LLG dynamics, is the computation of the torque in Eq.~(\ref{eq:torque}) due to itinerant electrons. Using Eq.~(\ref{eq:free-energy}) and (\ref{eq:E_elec1}) for the electron free-energy and internal energy, respectively, the derivative of $\mathcal{F}$ is given by
\[
	\frac{\partial \mathcal{F}}{\partial \mathbf S_i} = \sum_m f_m \frac{\partial \varepsilon_m}{\partial \mathbf S_i} + \sum_m \varepsilon_m \frac{\partial f_m}{\partial \mathbf S_m} - T \frac{\partial \mathcal{S}}{\partial \mathbf S_i}
\]
Using Eq.~(\ref{eq:S_elec}) for the electron entropy, it can be shown after some algebra that the last two terms in the above expression cancel each other. This can also be derived directly from the free-energy expression
\begin{eqnarray}
	\mathcal{F} = -k_B T\, \sum_m \left(1 + e^{-\beta \varepsilon_m }  \right).
\end{eqnarray}
Taking the derivative with respect to spin, we have
\begin{eqnarray}
	\label{eq:dF_dS}
	\frac{\partial \mathcal{F}}{\partial \mathbf S_i} &=& -k_B T \sum_m \frac{e^{-\beta \varepsilon_m}}{1 + e^{-\beta \varepsilon_m}} \frac{\partial \left(-\beta \varepsilon_m \right)}{\partial \mathbf S_m} \nonumber \\
	& = & \sum_m f_m \frac{\partial \varepsilon_m}{\partial \mathbf S_i}.
\end{eqnarray}
A similar result has been obtained previously in the context of quantum molecular dynamics~\cite{weinert92,wentzcovitch92}. To further simplify this expression, we write the electron Hamiltonian in terms of quasi-particle number operators: $\mathcal{H} = \sum_m \varepsilon_m \hat{n}_m$, where the eigen-energies depend on the instantaneous spin configuration $\varepsilon_m = \varepsilon_m(\{\mathbf S_i(t) \} )$. We thus have
\begin{eqnarray}
	\frac{\partial \hat{\mathcal{H}}}{\partial \mathbf S_i} = \sum_m \hat{n}_m \frac{\partial \varepsilon_m}{\partial \mathbf S_i}.
\end{eqnarray}
Next we average over the electrons and use the the identity $\langle \hat{n}_m \rangle = f_m$. We then obtain a generalization of the Hellmann-Feynman theorem
\begin{eqnarray}
	\label{eq:T_i_eq}
	\mathbf T_i = - \left\langle \frac{\partial \mathcal{H}}{\partial \mathbf S_i} \right\rangle = -\frac{1}{\mathcal{Z}_c} {\rm Tr}_c\left(e^{-\beta \mathcal{H}} \frac{\partial \mathcal{H}}{\partial \mathbf S_i} \right) ,
\end{eqnarray}
From the explicit form of the DE Hamiltonian in Eq.~(\ref{eq:H_DE}), we can further simplify the expression Eq.~(\ref{eq:T_i_eq}) [Note: repeated greek indices imply summation over spins]
\begin{eqnarray}
	\mathbf T_i(t) = J \bm\sigma^{\,}_{\alpha\beta} \, \rho^*_{i\beta, i\alpha}(t),
\end{eqnarray}
where we have introduced the quasi-equilibrium one-electron correlation function or density matrix
\begin{eqnarray}
	\rho^*_{i\alpha, j\beta}(t) \equiv  \langle \hat{c}^\dagger_{j\beta} \hat{c}^{\;}_{i \alpha} \rangle = \frac{1}{\mathcal{Z}_c} {\rm Tr}_c \left(e^{-\beta \mathcal{H}(\{\mathbf S_i(t)\})} \hat{c}^\dagger_{j\beta} \hat{c}^{\;}_{i \alpha} \right). \nonumber \\
\end{eqnarray}
The density matrix can be straightforwardly computed from the eigenvalues and eigenvectors of the Hamiltonian matrix Eq.~(\ref{eq:H_matrix}). However, exact diagonalization for $\rho^*_{i\alpha, j\beta}$ is not feasible for large-scale simulations, since its computational time scales cubically with the system size. Linear-scaling method for the calculation of density matrix can be achieved using, e.g. the kernel polynomial method (KPM) or recent machine learning techniques.

\subsection{Non-adiabatic Ehrenfest dynamics of double-exchange systems}

For non-adiabatic dynamics of the DE system, both electrons and spins evolve with time according to their respective governing dynamical equations. We first consider the case in which the electrons are in a pure state described by an instantaneous many-electron wave function $|\Psi(t) \rangle$. The evolution of this wave function is governed by the Schr\"odinger equation
\begin{eqnarray}
	\label{eq:Sch_eq}
	i \hbar \frac{d |\Psi \rangle}{dt} = \mathcal{H}\!\left(\{\mathbf S_i(t) \}\right) \, |\Psi \rangle.
\end{eqnarray}	
Here $\mathcal{H}$ is the DE Hamiltonian Eq.~(\ref{eq:H_DE}), which depends on time through the spins. 
To describe the spin dynamics, one then define an effective energy $\mathcal{E}(\Psi) \equiv \langle \Psi | \mathcal{H} | \Psi \rangle$, which is the energy corresponding to the electron wave function $|\Psi \rangle$. The torques acting on spins are, again, given by the Hellmann-Feynman formula
\begin{eqnarray}
	\mathbf T_i = -\frac{\partial \langle \Psi | \mathcal{H} | \Psi \rangle}{\partial \mathbf S_i} = -\langle \Psi | \frac{\partial \mathcal{H}}{\partial \mathbf S_i } |\Psi \rangle, 
\end{eqnarray}
Here we have neglected the derivatives $\partial |\Psi \rangle / \partial \mathbf S_i$ since the electron wave function $|\Psi\rangle$ does not explicitly depend on spins. Instead, as described above in Eq.~(\ref{eq:Sch_eq}), the electrons coupled to spins through a time-dependent Hamiltonian. For closed DE system, the energy conserving spin dynamics is described by the Landau-Lifshitz equation, which is given by Eq.~(\ref{eq:LLG}) with $\alpha = 0$:
\begin{eqnarray}
	\label{eq:LL_c}
	\frac{d\mathbf S_i}{dt} &=&  \mathbf S_i \times \langle \Psi | \frac{\partial \mathcal{H}}{\partial \mathbf S_i } |\Psi \rangle \\
	& =& - J \mathbf S_i \times \bm\sigma^{\,}_{\alpha\beta} \langle \Psi(t) | \hat{c}^\dagger_{i\alpha} \hat{c}^{\,}_{i \beta} |\Psi(t) \rangle \nonumber
\end{eqnarray}
These two coupled equations~(\ref{eq:Sch_eq}) and (\ref{eq:LL_c}) offer a complete description of the non-adiabatic dynamics for closed DE-type systems~\cite{koshibae09,koshibae11,ono17}, similar to the Ehrenfest dynamics for quantum MD simulations~\cite{marx09}.

Importantly, since the DE Hamiltonian is bilinear in fermion operators, the many-electron wave function $|\Psi(t) \rangle$ is a time-dependent single Slater determinant state~\cite{imada89}
\begin{eqnarray}
	|\Psi(t) \rangle = \prod_{m = 1}^{N_e} \hat{\psi}^\dagger_m(t) |0\rangle,
\end{eqnarray}
where the time-varying quasi-particle operators satisfy the Heisenberg equation of motion
\begin{eqnarray}
	\frac{d\hat{\psi}_m}{dt} = \frac{i}{\hbar} [\hat{\mathcal{H}}(\{\mathbf S_i(t) \}), \, \hat{\psi}_m(t) ],
\end{eqnarray}
and are subject to an initial condition such that $\{\hat{\psi}^\dagger_m(0)\}$ diagonalize the DE Hamiltonian at $t = 0$, i.e.
\begin{eqnarray}
	\hat{\mathcal{H}}\!\left(\{\mathbf S_i(0)\}\right) = \sum_m \varepsilon_m\, \hat{\psi}^\dagger_m(0) \hat{\psi}_m(0).
\end{eqnarray}
However, the numerical integration of the Slater determinant is rather cumbersome. So far, this method has been applied to study the photo-induced dynamics of rather small lattices~\cite{koshibae09,koshibae11,ono17}. 

\subsection{Landau-Lifshitz-von~Neumann dynamics for DE systems}

An alternative formulation of the non-adiabatic Ehrenfest dynamics is based on the single-particle density matrix. This approach also allows for inclusion of correlations in the initial state in the non-adiabatic electron dynamics. To this end, we switch to Heisenberg picture and consider time-dependent operators. We define an effective energy functional
\begin{eqnarray}
	\mathcal{E}(\{\mathbf S_i\}) &\equiv& {\rm Tr}\Bigl[\hat{\varrho}^{(0)} \hat{\mathcal{H}}(\{\mathbf S_i\}) \Bigr] \nonumber \\
	&\equiv& \sum_{\lambda} P_\lambda \langle \Phi_\lambda | \hat{\mathcal{H}}(\{\mathbf S_i\}) |\Phi_\lambda \rangle.
\end{eqnarray}
Here $\hat{\varrho}^{(0)} = \sum_\lambda P_\lambda |\Phi_\lambda \rangle \langle \Phi_\lambda |$ denotes an initial many-electron density matrix, and $\{|\Phi_\lambda\rangle \}$ form an ensemble of initial many-electron state. For the case of pure state, $\hat{\varrho}^{(0)} = |\Phi\rangle \langle \Phi|$, and when $|\Phi\rangle$ is given by the ground state $ |\Psi\rangle$, this energy is reduced to the case discussed above. For application to elementary excitations of a finite-temperature thermal state, we take $|\Phi_\lambda\rangle$ to be the many-electron eigenstates of the Hamiltonian at $t = 0$, and $P_\lambda \propto e^{-\beta \mathcal{E}_\lambda}$ to be the Boltzmann distribution. 

Given this energy functional, the torques acting on spins are computed as
\begin{eqnarray}
	\label{eq:T_rho}
	\mathbf T_i(t) &=& -\frac{\partial {\rm Tr}(\hat{\varrho}^{(0)}\,\hat{\mathcal{H}})}{\partial \mathbf S_i} \\
	&=& -{\rm Tr}\biggl( \hat{\varrho}^{(0)}\, \frac{\partial \hat{\mathcal{H}}}{\partial \mathbf S_i} \biggr) 
	= J \bm\sigma^{\,}_{\alpha\beta} \, \rho^{\,}_{i\beta, i\alpha}(t). \nonumber
\end{eqnarray}
Note that repeated greek indices imply summation over spins. 
Here we have also invoked the general Hellmann-Feynman theorem for the second identity. And in the third equality, we have used the explicit form of the Hund's rule coupling in DE Hamiltonian~(\ref{eq:H_DE}) and introduced the time-dependent single-electron density matrix
\begin{eqnarray}
	\rho_{i\alpha,j\beta}(t) \equiv {\rm Tr}\bigl[ \hat{\varrho}^{(0)} \, \hat{c}^\dagger_{j\beta}(t) \, \hat{c}^{\;}_{i \alpha}(t) \bigr].
\end{eqnarray}
Substitute Eq.~(\ref{eq:T_rho}) into the Landau-Lifshitz equation~(\ref{eq:LLG}) and set $\alpha = 0$, we obtain 
\begin{eqnarray}
	\label{eq:LL3}
	\frac{d\mathbf S_i}{dt} = -J \mathbf S_i \times \bm\sigma^{\,}_{\alpha\beta} \, \rho^{\,}_{i\beta, i\alpha}(t).
\end{eqnarray}
This spin dynamics equation has to be supplemented by the equation of motion for the density matrix $\rho_{i\alpha, j\beta}$, which can be obtained using the Heisenberg equation for the electron or fermion operators, e.g.
\begin{eqnarray}
	\frac{d\hat{c}_{i\alpha}}{dt} = \frac{1}{i\hbar} [\hat{c}_{i\alpha}, \hat{\mathcal{H}}] = \frac{1}{i\hbar} H^{\,}_{i\alpha, j\beta}\,\hat{c}^{\,}_{j\beta}.
\end{eqnarray}
The resultant equation is the von~Neumann equation with the one-particle Hamiltonian matrix $H_{i\alpha, j\beta}$, i.e. $d\rho/dt = [H, \rho]/i\hbar$. Or explicitly
\begin{eqnarray}
	\label{eq:vN}
	\frac{d\rho_{i\alpha, j\beta}}{dt} &=& i \sum_k (t_{ik} \, \rho_{k\alpha, j\beta} - \rho_{i\alpha, k\beta} \, t_{kj})  \\
	&+& i J (\mathbf S_i \cdot \bm \sigma_{\alpha\gamma} \, \rho_{i\gamma, j\beta} - \rho_{i\alpha, j\gamma} \, \bm\sigma_{\gamma\beta} \cdot \mathbf S_j). \nonumber
\end{eqnarray}
The coupled ordinary differential equations (\ref{eq:LL3}) and (\ref{eq:vN}), called the Landau-Lifshitz-von~Neumann dynamics, offer a complete description for the evolution of the DE system. Numerically, thanks to the sparsity of the Hamiltonian matrix, the von~Neumann Eq.~(\ref{eq:vN}) can be efficiently integrated using sparse matrix multiplication algorithms.

\newpage


\begin{thebibliography}{99}
\bibitem{note_tau} Here we have used $\tau = 187/91 \approx 2.054$ from the 2D percolation theory.
\end{thebibliography}
\end{document}